\documentstyle[12pt]{article}

\catcode`@=11
\def\@sim#1#2{\setbox0=\hbox{$\sim$}\lower.9\ht0\vbox{\baselineskip0pt
          \lineskip0.1ex\ialign{$\m@th#1\hfill##\hfill$\crcr#2\crcr
          \sim\crcr}}}
\def\lsim{\mathrel{\mathpalette\@sim<}}
\def\gsim{\mathrel{\mathpalette\@sim>}}
\catcode`@=12

\newcommand{\rsection}[1]{\setcounter{equation}{0}\section{#1}}

\newcommand{\be}{\begin{equation}}
\newcommand{\ee}{\end{equation}}
\newcommand{\bea}{\begin{eqnarray}}
\newcommand{\eea}{\end{eqnarray}}

\def\half{{1\over2}}

\def\fourth{{1\over4}}

\def\sixth{{1\over6}}
\def\d{{\rm d}}
\def\intdx{\int\d^d x\/}
\def\intdy{\int\d^d y\/}
\def\lag{{\cal L}}
\def\ln{{\rm ln\/}}

\def\Tr{{\rm Tr\/}}
\def\det{{\rm det\/}}
\def\order{{\cal O}}

\def\llambda{\lambda}
\def\g{g}

\def\cphi{\bar\phi}
\def\cA{\bar A}
\def\cF{\bar F}

\def\cU{U}

\def\cPsi{\bar\psi}
\def\cnabla{\bar\nabla}
\def\cf{f}
\def\n{\bar\nabla}

\def\fxi{\xi}
\def\feta{\eta}
\def\l{d}

\def\tildeD{{\tilde D}}

\def\varD{{\cal D}}
\def\bD{\bar D}
\def\bX{\bar X}

\def\bDelta{\bar \Delta}
\def\X{\vartheta}
\def\ttD{{\hat D}}

\def\pole{\Lambda_\varepsilon(2m^2)}
\def\poleMH{\Lambda_\varepsilon(M_H^2)}
\def\MZ{M_W}
\def\PT{PT}

\def\nf{\tilde f}
\def\nft{\hat f}
\def\ty{\hat y}
\def\tlag{{\hat\lag}}

\newcommand{\artref}[4]{{\rm #1}, {\it #2} {\bf #3} #4}

\newcommand{\bookref}[2]{{\rm #1}, #2}

\begin{document}

\begin{titlepage}

\def\mytoday#1{{ } \ifcase\month \or
 January\or February\or March\or April\or May\or June\or
 July\or August\or September\or October\or November\or December\fi
%\space\number\day ,
 \space \number\year}

\vspace*{-2.15cm}
\indent\hspace*{10cm}\mbox{BUTP-95/05}\newline
\indent\hspace*{10cm}\mbox{HUTP-95/A020}
\vspace*{.7cm}
\begin{center}
{\LARGE Effective Field Theory for a Heavy Higgs: a Manifestly Gauge Invariant
Approach}
\footnote{Work supported in part by Deutsche Forschungsgemeinschaft,
	by Schweizerischer Nationalfonds,
        and by NSF grant PHY-92-18167.} \\ [1.0cm]

   A. Nyf\/feler

\vspace*{0.2cm}
    Institute of Theoretical Physics \\
    University of Bern \\
    Sidlerstrasse 5, CH-3012 Bern, Switzerland \\[0.5cm]

   and \\[0.5cm]

   A. Schenk\footnote{New address after July 1st: Univ.\ Karlsruhe, Theor.\
Teilchenphysik, Kaiserstr.\ 12, D-76128 Karlsruhe, Germany}

\vspace*{0.2cm}
    The Physics Laboratories\\
    Harvard University \\
    Cambridge, MA 02138, USA \\[1cm]

 June 1995
\vspace*{.5cm}

\begin{abstract}
For large values of the Higgs boson mass the low energy structure of the gauged
linear sigma model in the spontaneously broken phase can adequately be
described by an effective field theory. In this work we present a manifestly
gauge invariant technique to explicitly evaluate the
corresponding effective Lagrangian from the underlying theory. In order to
demonstrate the application of this functional method, the effective field
theory of the abelian Higgs model is thoroughly analyzed. We stress that this
technique does not rely on any particular property of the abelian case. The
application to the non-abelian theory is outlined.
\end{abstract}
\end{center}
\end{titlepage}

\rsection{Introduction}

The method of effective field theory has repeatedly been used in the analysis
of the symmetry breaking sector of the Standard Model \cite{L_eff_SM}. It
provides a convenient and model independent parametrization of various
scenarios discussed in the literature, such as a heavy Higgs~\cite{heavy_Higgs}
or technicolor models~\cite{technicolor}, regarding the nature of the
spontaneous breaking of the electroweak symmetry. In this approach, the unknown
physics is hidden in the low energy constants of an effective Lagrangian, which
describes the effective field theory. Thus, in order to analyze the
phenomenological implications of precision experiments with presently
accessible energies on physics beyond the Standard Model, it is necessary to
study the relationships between the constants of this Lagrangian and the
parameters of a given underlying theory.

The physics in the low energy region of a full theory is adequately described
by an effective field theory if corresponding Green's functions in both
theories have the same low energy structure. One can take this matching
requirement as the definition of the effective field theory. It determines
functional relationships between the low energy constants of the effective
Lagrangian and the parameters of the underlying theory. Furthermore, if the
coupling constants are weak in the low energy region, perturbative methods can
be used to explicitly evaluate these relationships.

Functional techniques to evaluate the effective Lagrangian for a given
underlying theory have recently been thoroughly discussed~\cite{LSM}. However,
their application to gauge theories is not straightforward. The present article
will continue the discussion and give a thorough account of this problem.

The special r\^ole of gauge theories is readily understood. The effective field
theory analysis of the symmetry-breaking sector of the Standard Model should
not make any particular assumptions about the underlying theory -- apart from
symmetry properties and the existence of a mass gap. Conceptually, this also
requires parametrizing low energy phenomenology by a gauge invariant effective
Lagrangian. The definition of Green's functions, on the other hand, usually
does not reflect the symmetry properties of a gauge theory. Gauge invariance is
broken, and the off-shell behaviour of Green's functions is gauge dependent.

An effective field theory defined as in the second paragraph will correctly
describe low energy physics. However, if Green's functions which enter the
matching relations do not reflect the symmetry properties of the full theory,
it will also include the corresponding gauge artifacts. It is not clear at all,
whether such an effective field theory can be described by an effective
Lagrangian which is gauge invariant.

Without resolving that issue, this approach was taken in Ref.~\cite{H_M} in
order to determine the low energy constants at order~$p^4$ in the case, where
the Higgs mass in the Standard Model is heavy. Under the additional assumption
that the scalar coupling~$\llambda$ is not too strong, perturbative methods
can indeed be used to evaluate the corresponding effective Lagrangian. To pin
down the low energy constants at order~$p^4$, the authors matched 1PI Green's
functions of the gauge fields. The off-shell behaviour of these functions in
both the full and the effective theory is gauge dependent. Thus, fine tuning
of the gauge dependent structure is necessary in order to avoid
inconsistencies. For the case at hand, it was sufficient to choose appropriate
gauge fixing conditions in both theories. However, these Green's functions
involve gauge dependent terms of arbitrary order in the low energy expansion.
Furthermore, it was shown in Ref.~\cite{E_M} that the 1PI Green's functions of
the gauge fields depend on the parametrization of the Goldstone boson fields.
To account for this effect, the matching relations should involve all
light-particle Green's functions, including those of the Goldstone bosons. It
is rather doubtful that it will be possible to maintain the off-shell matching
of all these Green's functions, yet still describe the effective field theory
defined that way by a gauge invariant effective Lagrangian and a fine tuned
gauge fixing term. We think that this approach will generally involve an
effective Lagrangian which is not gauge invariant.

In order to avoid these problems, the authors of Ref.~\cite{E_M} suggested that
all gauge dependence be eliminated from the matching relations. However,
the approach they chose does not work in general. They projected onto the
transverse components of the gauge fields in connected Green's functions. In a
non-abelian gauge theory, this does not yield gauge invariant quantities. For
the case of a heavy Standard Model Higgs, it turned out that the matching
relations between projected Green's functions at order~$p^4$ do not depend on
the gauge fixing parameter~$\xi$. However, independence of this parameter is
only a necessary condition for gauge invariance. Since projecting onto the
transverse components of Green's functions does not yield gauge invariant
quantities in the non-abelian case, this method is inherently gauge
dependent.

Yet another approach was taken in Ref.~\cite{D_GK}, in which the background
field action of both the full and the effective theory are matched. Here, too,
the off-shell behaviour of Green's functions enters the matching relation. One
particular property of this action is its gauge invariance with respect to
gauge transformations of the background fields, which is achieved by an
appropriate choice of the gauge fixing term. This may be quite useful in
certain applications, but it does not solve the problems described above.
Though gauge invariant,  the background field action is not gauge independent.
The 1PI Green's functions explicitly depend on the gauge fixing term, which was
introduced in the definition of the background field action. Thus, it is again
necessary to fine tune the gauge fixing terms in both the full and the
effective theory order by order in the low energy expansion. Furthermore, one
still has to make sure that no gauge dependence enters the effective
Lagrangian. Without further proof, independence of the gauge fixing parameter
should only be considered as a necessary condition. In this work we will not
distinguish between gauge invariance and gauge independence. The phrase gauge
invariance will include both meanings.

This discussion indicates that any approach to determine the effective
Lagrangian for a given underlying theory should match only gauge invariant
quantities. Then one does not have to worry about any gauge artifact which
otherwise might enter the effective field theory. Perhaps the most
straightforward idea that comes to mind is to match only $S$-matrix elements.
However, this approach is quite cumbersome, and one would rather like to use
functional techniques like those described in Ref.~\cite{LSM}. The main purpose
of this article is to formulate these techniques for the case of gauge theories
under the condition that gauge invariance is manifest throughout the
calculation. In order to avoid technical difficulties we have chosen the
abelian Higgs model as a simple example to demonstrate this approach. We
emphasize that our analysis does not rely on any particular property of the
abelian theory. In fact, we will outline the application to the non-abelian
case in the last Section of this article.

In the next Section we will briefly introduce the abelian Higgs model and
discuss a manifestly gauge invariant technique to evaluate the generating
functional of Green's functions of gauge invariant operators. In Section 3 the
low energy constants which describe the low energy structure of the one-loop
approximation in the abelian Higgs model up to order~$p^6$ are determined. In
Section 4 we will discuss the evaluation of 2- and higher loop corrections in
our approach.  Section 5 is devoted to renormalization. Finally, our results
are summarized and discussed in Section 6.

\rsection{The Abelian Higgs Model to One Loop}

The Lagrangian of the abelian Higgs Model is given by
\begin{equation} \label{lag0}
	\lag = \half\nabla_\mu\phi^T\nabla_\mu\phi - \half m^2 \phi^T\phi
		+ {\llambda\over4} (\phi^T\phi)^2
		+ {1\over4\g^2} F_{\mu\nu} F_{\mu\nu} \ ,
\end{equation}
where $\phi^A$ is a 2-component scalar field, coupled to the gauge field
$A_\mu$ through the covariant derivative
\begin{equation} \label{cov_deriv}
	\nabla_\mu\phi = \partial_\mu\phi + A_\mu T_e \phi \ .
\end{equation}
The generator of the $O(2)$ symmetry is of the form
\begin{equation}
	T_e =  \left( \begin{array}{rr}
           	0  & 1 \\
		-1 & 0
\end{array}\right) \ .
\end{equation}
The field strength is given by the expression $F_{\mu\nu} = \partial_\mu A_\nu
- \partial_\nu A_\mu$.  The Lagrangian~(\ref{lag0}) is invariant under gauge
transformations of the form
\begin{eqnarray} \label{gaugetrafo1}
	\phi \rightarrow \exp{( \omega T_e ) } \phi \\
	A_\mu \rightarrow A_\mu - \partial_\mu \omega \label{gaugetrafo2} \ .
\end{eqnarray}
For computational convenience we are working in euclidean space-time.

For $m^2 > 0$ the classical potential has its minimum at a nonzero value
$\phi^T\phi = m^2/\llambda$ and the $O(2)$ symmetry is spontaneously broken.
Accordingly, the 2-com\-po\-nent field $\phi^A$ describes one massive mode,
the Higgs particle, and one Goldstone boson. Thus, the spectrum of the abelian
Higgs model consists of two massive states: the spin~$0$ Higgs particle, and
the spin~$1$ gauge boson.

In order to have nontrivial solutions of the equations of motion, we
furthermore have to couple external sources to the gauge boson and the Higgs
field. The appropriate choice for these sources is crucial for an analysis
which should be manifestly gauge invariant. Suppose, we couple
the field $\phi^A$ to a set of external fields with spin $0$ by way of the
source term
\begin{equation} \label{gbs1}
		f^T\phi \ .
\end{equation}
If one considers both the gauge field $A_\mu$ as well as the scalar field
$f^A$ as external, one recovers exactly the situation of the ungauged linear
sigma model~\cite{LSM}. In that case, the generating functional was defined
to be
\begin{equation} \label{uggenfunc}
	e^{-W^{ug}_\sigma[A_\mu, f]}
	= \int \d\mu[\phi] e^{-\intdx \left( \lag - f^T \phi \right)} \ .
\end{equation}
Derivatives of this functional with respect to the spin $0$ field $f^A$
generate Green's functions of the scalar fields $\phi^A$, while derivatives
with respect to the spin $1$ field $A_\mu$ generate Green's functions of the
current $(\partial_\mu\phi)^T T_e \phi$. Furthermore, the generating
functional turns out to be gauge invariant under gauge transformations of the
form~(\ref{gaugetrafo1}) and~(\ref{gaugetrafo2}), which act on the external
sources. This property summarizes the Ward identities of the ungauged linear
sigma model.

However, the situation is different if the gauge fields are dynamical
degrees of freedom. In this case, only the field $f^A$ is an external source.
The equations of motion are determined by the condition that the classical
action be stationary, i.e.,
\begin{equation}
	\delta\intdx \left( \lag - f^T\phi \right) = 0 \ .
\end{equation}
The problem associated with this source term becomes obvious if variations of
the fields corresponding to infinitesimal gauge transformations are considered.
Since the Lagrangian itself is gauge invariant, we end up with the condition
\begin{equation}  \label{gbc1}
	f^T T_e \phi = 0 \ .
\end{equation}
Thus, the equations of motion have a solution only under the constraint that
the gauge degree of freedom of the scalar field $\phi^A$ is frozen in a very
particular way, depending on the direction of the external source $f^A$.  If
an external current $j_\mu$ is coupled to the gauge field via
\begin{equation}   \label{gbs2}
	j_\mu A_\mu \ ,
\end{equation}
the condition analogous to Eq.~(\ref{gbc1}) is given by
\begin{equation}
	\partial_\mu j_\mu = 0 \ ;
\end{equation}
i.e., the external current has to be conserved.

Note that the source term~(\ref{gbs1}) explicitly breaks the symmetry in both
the gauged and the ungauged linear sigma model. In the latter case this does
not hurt. The generating functional~(\ref{uggenfunc}) still reflects the
symmetry properties of the ungauged theory. Although fluctuations corresponding
to global~$O(2)$ transformations also yield a constraint, it does not affect
the equations of motion. In fact, it is trivially satisfied by the classical
fields.

If the symmetry is gauged, however, a symmetry breaking external source
couples to the gauge degrees of freedom.  This does not seem to be a suitable
perturbation of the system. The action does not have a stationary point, if
such a source term is present. Thus, we are naturally led to consider only
external sources which couple to gauge invariant operators.  Green's functions
of these operators are the proper objects to analyze in a gauge theory.

We will see below that a manifestly gauge invariant treatment is
indeed possible if the analysis involves only gauge invariant sources.
This is not only true at the classical level, but also if quantum corrections
are taken into account.

For our analysis, a convenient choice of gauge invariant operators is the
scalar density $\phi^T\phi$ and the field strength $F_{\mu\nu}$:
\begin{eqnarray} \label{full_lag}
\lag_\sigma & = & \half\nabla_\mu\phi^T\nabla_\mu\phi - \half m^2 \phi^T\phi
		+ {\llambda\over4} (\phi^T\phi)^2
		+ {1\over4\g^2} F_{\mu\nu} F_{\mu\nu} \nonumber \\
& & 		- \half h \phi^T\phi - \half k_{\mu\nu} F_{\mu\nu}
		- c_{hh} h^2 - c_{mh} m^2 h \ .
\end{eqnarray}
The generating functional $W_\sigma[h,k_{\mu\nu}]$ is defined by a path
integral
\begin{equation} \label{genfunc}
	e^{-W_\sigma[h,k_{\mu\nu}]} = \int \d\mu[\phi,A] e^{-\intdx
		\lag_\sigma} \ .
\end{equation}
Derivatives of this functional with respect to the field $h$ generate Green's
functions of the scalar density $\phi^T\phi$, while derivatives with respect to
the source $k_{\mu\nu}$ generate Green's functions of the field strength
$F_{\mu\nu}$. In the spontaneously broken phase, these Green's functions have
one-particle poles from both bosons. Thus, one can even extract $S$-matrix
elements from the generating functional~(\ref{genfunc}).

Green's functions of the composite operator $\phi^T\phi$ are more singular at
short distances than Green's functions of the scalar field itself. Time
ordering of these operators gives rise to ambiguities and corresponding Green's
functions are only unique up to contact terms. We will see below that
renormalization of the two constants $c_{hh}$ and $c_{mh}$ in the
Lagrangian~(\ref{full_lag}) is necessary to render the generating
functional~(\ref{genfunc}) finite after the regulator is removed. The ambiguity
is then reflected by the presence of the finite parts of these two constants.

At tree-level, the generating functional is given by
\begin{equation}
	W_\sigma[h,k_{\mu\nu}]
		= \intdx \lag_\sigma(\cphi, \cA_\mu) \ ,
\end{equation}
where $\cphi$ and $\cA_\mu$ are determined by the equations of motion
\begin{eqnarray} \label{eqmo1phi}
	( -\cnabla_\mu\cnabla_\mu
		- m^2 + \llambda \cphi^T\cphi
		- h ) \cphi & = & 0 \\
	(-\delta_{\mu\nu} \Box + \partial_\mu\partial_\nu) \cA_\mu
		+ g^2 ( J_\nu
		+ \partial_\mu k_{\mu\nu} ) & = & 0 \ ,    \label{eqmo1A}
\end{eqnarray}
with
\begin{equation}
	\cnabla_\mu = \partial_\mu + T_e \cA_\mu \ .
\end{equation}
The $O(2)$ current $J_\mu$ is of the form
\begin{equation}
	J_\mu  \doteq (\cnabla_\mu \cphi)^T T_e \cphi \ .
\end{equation}
It is useful to introduce the following parametrization of the classical
solution $\cphi^A$:
\begin{equation}
	\cphi^A = {m\over\sqrt{\llambda}} R U^A \  , \qquad U^TU = 1 \ ,
\end{equation}
where the radial variable R describes the massive mode, while the field $U^A$
corresponds to the Goldstone boson. In terms of the new variables,
Eqs.~(\ref{eqmo1phi}) and (\ref{eqmo1A}) are of the form:
\begin{eqnarray} \label{eqmo2R}
  (-\Box + m^2 ( R^2 - 1 ) + \cnabla_\mu U^T  \cnabla_\mu U - h ) R &=& 0 \\
	\partial_\mu J_\mu  &= & 0 \label{eqmo2cc} \\
	(-\delta_{\mu\nu} \Box + \partial_\mu\partial_\nu) \cA_\mu
		+ g^2 (J_\nu
		+ \partial_\mu k_{\mu\nu} ) & = & 0 \ . \label{eqmo2A}
\end{eqnarray}
Several things about these equations are worth being noticed. Gauge invariance
implies that they have a whole class of solutions. Every two representatives
are related to each other by a gauge transformation. Due to Eq.~(\ref{eqmo2cc})
and the asymmetry of the external source $k_{\mu\nu}$, the vector field couples
to a conserved current. In order to solve Eqs.~(\ref{eqmo2cc})
and~(\ref{eqmo2A}) we write the Goldstone boson field as
\begin{equation}
	U = e^{w T_e } U_0 \ ,
\end{equation}
where~$U_0$ is an arbitrary but constant~$O(2)$ vector. The equations of
motion~(\ref{eqmo2R})-(\ref{eqmo2A}) then involve the three quantities~$R,
\cA^T_\mu$, and~$\cA^L_\mu + \partial_\mu w$, where~$\cA_\mu^L =  \partial_\mu
({1/\Box}) \partial_\nu\cA_\nu$ and~$\cA_\mu^T = \cA_\mu - \cA_\mu^L$. Neither
the equations of motion nor the action depend on the vector~$U_0$. The radial
component~$R$ describes the Higgs boson, while the transverse and the
longitudinal degrees of freedom of the gauge boson are described by the two
gauge invariant fields~$\cA^T_\mu$ and~$\cA_\mu^L + \partial_\mu w$. Picking a
certain representative for our solution is equivalent to specifying how the
longitudinal degree of freedom is split between the two fields~$\cA_\mu^L$
and~$U^A$. Two extreme choices are the unitary gauge,~$w = \mbox{const}$, and
the condition~$\partial_\mu \cA_\mu = 0$.

The one-loop contribution to the generating functional can be evaluated with
the saddle point method. Using the parametrization
\begin{eqnarray}
	\phi &=& \cphi + f 		\label{fluct1phi} \\
	A_\mu &=& \cA_\mu + g q_\mu  	\label{fluct1A}
\end{eqnarray}
for the fluctuations around the classical fields $\cphi^A$ and $\cA_\mu$, one
obtains
\begin{equation} \label{W_pathint}
 e^{-W_\sigma[h,k_{\mu\nu}]} =
		e^{-\intdx\lag_\sigma(\cphi,\cA_\mu)}
		\int\d\mu[\phi,A] e^{- \half \intdx y^T \tildeD y} \ ,
\end{equation}
with
\begin{equation}
	y = \left( \begin{array}{c}
		f \\
		q_\mu
\end{array}\right)   \ ,
\end{equation}
and
\begin{equation}
	\tildeD =  \left( \begin{array}{cc}
                D^{ff} & D^{fq} \\
		D^{qf} & D^{qq}
\end{array}\right)  \ .
\end{equation}
The components of the symmetric operator $\tildeD$ are given by
\begin{eqnarray}
	D^{ff} & = & -\cnabla_\mu\cnabla_\mu
			- ( m^2 - \llambda \cphi^T \cphi + h ){\rm\bf 1}
			+ 2 \llambda \cphi \cphi^T \\
	D^{fq}_\mu & = & - g T_e \cphi \partial_\mu
			 - 2 g T_e( \cnabla_\mu \cphi) \\
	D^{qf}_\nu & = & - g \cphi^T T_e \cnabla_\nu
			+ g (\cnabla_\nu\cphi)^T T_e \\
	D^{qq}_{\mu\nu} & = & -\delta_{\mu\nu}\Box + \partial_\mu \partial_\nu
				+ \delta_{\mu\nu} g^2 \cphi^T\cphi	\ .
\end{eqnarray}

Gauge invariance implies that the operator $\tildeD$ has zero eigenvalues
corresponding to fluctuations $y$ which are equivalent to infinitesimal gauge
transformations. Indeed, if $\cPsi^i \doteq (\cphi^A, \cA_\mu)$ is a
solution of the equation of motion, i.e., a stationary point of the classical
action,
\begin{equation} \label{geqmo}
	\left.{\delta S_\sigma\over\delta\psi^i}\right|_{\psi=\cPsi} = 0 \ ,
\end{equation}
then any gauge transformation~(\ref{gaugetrafo1}), (\ref{gaugetrafo2}) yields
another equivalent solution. Thus, differentiating
equation~(\ref{geqmo}) with respect to the gauge parameter $\omega$ one obtains
\begin{equation} \label{zero1}
	\left.
	{\delta^2 S_\sigma\over\delta\psi^i\delta\psi^j}
	{\delta\psi^j\over\delta\omega}
	\right|_{\psi=\cPsi} = 0 \ .
\end{equation}
This can also be verified explicitly, using the following representation of
the zero eigenvectors $\zeta$ in terms of scalar fields $\alpha$,
\begin{equation}
	\zeta = P\alpha \ ,
\end{equation}
where
\begin{equation}
	P = \left(\begin{array}{c}
               g T_e \cphi^A \\
		-\partial_\mu
\end{array}\right)   \ .
\end{equation}
Then Eq.~(\ref{zero1}) translates to the identity
\begin{equation}
  P^T \tildeD  =  \tildeD P  = 0 \ .
\end{equation}
Let $\alpha_m$ be the eigenvectors of the operator $P^T P$, i.e.
\begin{equation}
	P^T P \alpha_m = l_m \alpha_m \ .
\end{equation}
Then, the expansion of the fluctuation $y$ in terms of eigenvectors of the
operator $\tildeD$ is given by
\begin{equation}
	y = \sum_n a_n \xi_n + \sum_m b_m \zeta_m \ ,
\end{equation}
where $\zeta_m = P \alpha_m$ and $\xi_n$ have zero and non-zero eigenvalues,
respectively.

In order to evaluate the path integral~(\ref{W_pathint}), we use Polyakov's
method~\cite{Polyakov} and equip the space of fields with a metric
\begin{eqnarray}
    ||y||^2 & = & \intdx y^T y     \\
	    & = & \sum_n a_n^2 + \sum_m b_m^2 l_m \ .
\end{eqnarray}
With our choice for the scalar fields $\alpha_m$, the metric on the kernel of
the differential operator $\tildeD$ is indeed diagonal:
\begin{equation}
  g_{\bar m m} = \intdx \alpha_{\bar m} P^T P \alpha_m = \delta_{\bar m m}
	l_m\ .
\end{equation}
The volume element associated with this metric is then given by
\begin{equation}
\d\mu[\phi,A] = {\cal N} \prod_n \d a_n \prod_m \d b_m \sqrt{ \det P^T P } \ .
\end{equation}
The integration over the zero-modes yields the volume factor of the gauge
group, which can be absorbed by the normalization of the integral. The
remaining integral over the non-zero modes is damped by the usual gaussian
factor.  One obtains the following result for the one-loop generating
functional
\begin{equation} \label{1loopgf}
	W_\sigma[h,k_{\mu\nu}] = \intdx \lag_\sigma
				+ \half \ln\det^\prime\tildeD
				-\half \ln\det P^TP \ ,
\end{equation}
where $\det^\prime\tildeD$ is defined as the product of all non-zero
eigenvalues of the operator $\tildeD$. The evaluation of path integrals like
the one in Eq.~(\ref{W_pathint}), with a semi-definite quadratic form in the
exponent, is also discussed in the context of instanton
calculations~\cite{Corrigan_Goddard}.

Since zero and non-zero eigenvectors are orthogonal to each other, implying
$P^T\xi_n = 0$, one furthermore verifies the useful identity
\begin{equation} \label{usefullid}
%	\det( \tildeD + P P^T ) = \det^\prime \tildeD \; \det P^T P \ .
	\det^\prime \tildeD	= {\det( \tildeD + P P^T )\over \det P^T P}\ .
\end{equation}

Finally, by changing to a new basis, we separate the massless fluctuations
around the classical solution $\cphi^A$ which are bound to the sphere $U^T U =
1$, from the massive fluctuation along $U^A$:
\begin{equation}
	f = \fxi U + \feta \epsilon \ , \qquad \epsilon = T_e U \ .
\end{equation}
In order to describe the action of the differential operator $\tildeD$ in the
new basis, we introduce the following notation:
\begin{eqnarray} \label{doperatorsfirst}
	D	& = &  -\Box + m^2 ( R^2 - 1) + \cnabla_\mu U^T\cnabla_\mu U
							- h	\\
	d 	& = &   D + 2 m^2 R^2 \\
	\delta 	& = &   -2f_\mu \partial_\mu - (\partial_\mu f_\mu) \\
	\delta_\mu & = & - 2 M R f_\mu \\
	\Delta_\mu & = & - M R \partial_\mu - 2 M (\partial_\mu R)
	\label{doperatorsllast} \\
	f_\mu   & = & 	U^T \cnabla_\mu \epsilon \label{doperatorslast} \\
	\Theta 	& = & D - \delta^T d^{-1} \delta \\
	\X_\mu 	& = & \Delta_\mu - \delta^T d^{-1} \delta_\mu \\
	\varD_{\mu\nu} & = & D^{qq}_{\mu\nu} - \delta_\mu^T d^{-1} \delta_\nu
			  - \X_\mu^T \Theta^{-1} \X_\nu \ ,
	\label{doperatorslllast}
\end{eqnarray}
where
\begin{equation} \label{M_bare}
	M = m { \g\over\sqrt{\llambda}}
\end{equation}
is the bare mass of the gauge field.
We furthermore diagonalize the quadratic form $y^T \tildeD y$
by the transformation
\begin{eqnarray} \label{diagtrafo}
	\fxi 	& \to &  \fxi - d^{-1} \delta \feta
			      + d^{-1} \delta \Theta^{-1} \X_\mu q_\mu
			      - d^{-1} \delta_\mu q_\mu \nonumber \\
	\feta 	& \to &  \feta - \Theta^{-1} \X_\mu q_\mu \ .
\end{eqnarray}
The quadratic form in Eq.~(\ref{W_pathint}) now reads:
\begin{equation}
	\intdx y^T\tildeD y =  \intdx \left( \fxi d\fxi + \feta\Theta\feta
		+ q \varD q \right) \ .
\end{equation}
The transformation of the basis also changes the metric on the space of
functions. The zero-modes are now of the form
\begin{equation} \label{simplezeromode}
	\hat P \alpha_m
\end{equation}
with
\begin{equation}
 	\hat P^T = ( 0, 0, \partial_\mu) \ ,
\end{equation}
i.e., they correspond to longitudinal fluctuations
of the gauge field. The generating functional turns out to be
\begin{eqnarray} \label{1loopgfdiag}
	W_\sigma[h,k_{\mu\nu}] & = & \intdx \lag_\sigma
				  + \half \ln\det d + \half \ln\det \Theta
		        	  + \half \ln\det^\prime \varD \nonumber \\
		 	  & & 	  - \half \ln\det \hat P^T \hat P \ .
\end{eqnarray}
The last term is constant and will be absorbed in the overall normalization.

To further simplify the calculation, one can make use of the transversality of
the operator~$\varD$ and write
\begin{equation} \label{simplon}
	\varD_{\mu\nu} = \PT_{\mu\rho} \varD_{\rho\sigma} \PT_{\sigma\nu} \ ,
\end{equation}
where
\begin{equation}
	\PT_{\mu\nu} =
		\delta_{\mu\nu} - \partial_\mu {1 \over \Box} \partial_\nu
\end{equation}
projects on the non-zero modes. Eq.~(\ref{simplon}) implicitly involves the
equations of motion and indeed simplifies the evaluation of the determinant.
Using an identity similar to~(\ref{usefullid}), one obtains
\begin{equation}
	\ln\det^{\prime} \varD = - \ln\det d_M +
		\ln\det \left( d_M + \PT \sigma \PT \right) \ ,
\end{equation}
with
\begin{eqnarray}
	d_M 		& = & - \Box + M^2 \label{d_M} \\
        \sigma_{\mu\nu} & = & \delta_{\mu\nu} M^2 (R^2 - 1)
                                - \delta^T_\mu d^{-1} \delta_\nu
                                - \X^T_\mu \Theta^{-1} \X_\nu  \ .
				\label{sigma}
\end{eqnarray}

\rsection{The Effective Lagrangian}

In this section we will discuss the effective theory of the abelian Higgs model
for the case of a heavy Higgs boson mass, i.e.,
\begin{equation}
	p^2, M^2 \ll 2 m^2 \ .
\end{equation}
It is well known that the effective field theory can be described by an
effective Lagrangian. In our case this Lagrangian is gauge invariant, and
depends on the 2-component Goldstone boson field~$U^A$, confined to the
sphere~$U^T U = 1$, the vector field~$A_\mu$, and the external sources
\begin{equation}
 \lag_{eff} = \lag_{eff}\left( F_{\mu\nu}, U, \nabla_\mu U,
   \nabla_\mu\nabla_\nu U, h, k_{\mu\nu}, \ldots \right) \ .
\end{equation}
It describes the dynamics of one massive gauge boson. The corresponding
generating functional is defined as a path integral,
\begin{equation}
	e^{-W_{eff}[h,k_{\mu\nu}]}
	= \int \d\mu[U,A] e^{-\intdx \lag_{eff}} \ .
\end{equation}
The effective Lagrangian is determined by a matching relation, which requires
that both the full and the effective theory yield the same Green's
functions in the low energy region:
\begin{equation}  \label{match1}
	W_{\sigma}[h, k_{\mu\nu}] = W_{eff}[h, k_{\mu\nu}] \ .
\end{equation}
At low energies, these Green's functions have non-local contributions related
to the singularities of the light vector boson. These contributions drop out
of the matching relation and the remaining terms involve powers of the
coupling constants, the momenta and the mass. In order to systematically
evaluate the effective Lagrangian, we need to understand the counting of loops
in the full theory and of the low energy expansion in the effective theory.

In the ungauged linear sigma model, the loop expansion generates a power series
in the coupling constant~$\llambda$, while the low energy expansion at a given
loop-level produces powers of the momenta. Thus, it may seem that one has to
keep track of four quantities if the theory is gauged, i.e., powers of the
couplings~$\llambda$ and~$\g$ as well as powers of the momenta and the gauge
boson mass. However, such a counting scheme is ambiguous, since the gauge boson
mass itself depends on the coupling constants, as shown in Eq.~(\ref{M_bare}).
This expression also indicates that it will not be very transparent to count
mass factors in terms of the quantities~$\llambda$ and~$g^2$. The loop
expansion in the full theory generates positive powers of the
coupling~$\llambda$, while the low energy expansion produces negative powers
thereof. To simplify the bookkeeping, one may introduce a loop factor~$l$,
counting the number of loops in the full theory, and treat~$\llambda$ and~$g^2$
as order~$l$. In this case $n$-loop contributions will be of order~$l^{n-1}$.

Equivalently, one may discard the coupling constant~$g$ from the counting
scheme. This is a consequence of the definition of the vector field~$A_\mu$ in
Eq.~(\ref{cov_deriv}), which is scaled such that the constant~$\g$ does not
explicitly occur in the covariant derivative. As a result, this constant
naturally enters all loop correction only through the gauge boson mass~$M$.
Regarding the one-loop contributions to the generating functional, this can
readily be inferred from
Eqs.~(\ref{doperatorsfirst})-(\ref{doperatorslast}). As we will see in the
next Section, this is also true for all higher loops.  Furthermore, with this
bookkeeping powers of~$\llambda$ count the number of loops in the full theory.

In order to evaluate the low energy expansion at a given loop-level, we treat
the covariant derivative~$\nabla_\mu$, the gauge boson mass~$M$, and the
momenta as quantities of order~$p$, while the external source~$h$ is of
order~$p^2$.  The scalar fields, the Higgs mass $\sqrt{2} m$, and the external
source~$k_{\mu\nu}$ are quantities of order one. In counting the mass~$M$ as
order~$p$, the low energy expansion is carried out at a fixed ratio~$p^2/M^2$,
and correctly reproduces all singularities associated with the gauge boson.

The coherence of these rules requires that the coupling constant~$\g$ is
treated as a quantity of order~$p$. Note that this is different from the
usual dimensional analysis: the constant~$\g$ has dimension $(\mbox{mass})^0$,
yet it counts as order~$p$ in the low energy expansion. This is similar to
chiral perturbation theory in QCD~\cite{ChPT}, where, for example, the light
quark masses with dimension $(\mbox{mass})^1$ are treated as quantities of
order~$p^2$.

The effective Lagrangian is a sum of terms with an increasing number of
derivatives, mass factors and powers of the external fields,
\begin{equation} \label{Lageff}
	\lag_{eff} = \lag_2 + \lag_4 + \lag_6 + \ldots \ ,
\end{equation}
where $\lag_i$ is of order $p^i$. Furthermore, if the coupling of the scalar
field is not too strong, the low energy constants admit an expansion in powers
of the parameter~$\llambda$,
\begin{equation}  \label{lowconstexp}
	\l_i = {1 \over \llambda} \l_i^{tree}
		+ \l_i^{1-loop}
		+ \llambda \l_i^{2-loop} + \ldots \ ,
\end{equation}
corresponding to the loop expansion in the full theory. In this case the
accuracy of the effective field theory description is controlled by the order
of both the momentum and the coupling constant~$\llambda$. For values
of~$\llambda$ close to the strong coupling region, one may consider higher
orders in the expansion~(\ref{lowconstexp}). Large values of the momentum or
the gauge boson mass may require including higher orders in
Eq.~(\ref{Lageff}). In the following, we will determine this Lagrangian up to
order~$p^6$, and the low-energy constants up to order~$\llambda^0$.

In order to evaluate the low-energy constants, one can write down the most
general effective Lagrangian up to order~$p^6$, calculate the generating
functional in both the full and the effective theory, and solve the matching
relation~(\ref{match1}). However, we will make use of the fact that powers of
the constant~$\llambda$ count the number of loops. To evaluate the one-loop
contribution to the generating functional of the effective theory up to
order~$\llambda^0$, only the parameters~$\l_i^{tree}$ in
Eq.~(\ref{lowconstexp}) are relevant. They can be read off from the low energy
expansion of the classical action of the full theory, i.e., from

\begin{eqnarray} \label{clactR}
 \lefteqn{  \intdx\lag_\sigma(\cphi, \cA_\mu)  } \nonumber\\
       & = & \intdx \left(- {m^4\over 4\llambda} R^4
		+ {1\over4\g^2} \cF_{\mu\nu} \cF_{\mu\nu}
		- \half k_{\mu\nu} \cF_{\mu\nu}
		- c_{hh} h^2 - c_{mh} m^2 h \right)\ .
\end{eqnarray}
For slowly varying external fields, the behaviour of the massive mode~$R$ is
under control and the equation of motion~(\ref{eqmo2R}) can be solved
algebraically. The result is a series of local terms with increasing order
in~$p^2$. Since Eq.~(\ref{eqmo2R}) does not involve the parameter~$\llambda$,
all terms in the low energy expansion of the classical action count as
order~$\llambda^{-1}$, except for the two contact terms and the source term
for the field strength.  Note that the kinetic term of the vector field also
counts as order~$\llambda^{-1}$. In order to maintain a coherent scheme in the
presence of external sources, we treat the quantities~$c_{mh}, c_{hh}$
and~$k_{\mu\nu}$ as order~$\llambda^{-1}$. One obtains
\begin{eqnarray}
	\lag_2^{tree} & = & {m^2 \over 2 \llambda} \n_\mu \cU^T\n_\mu \cU
		       	+ {1\over4\g^2} \cF_{\mu\nu} \cF_{\mu\nu}
			- {m^2 \over 2 \llambda} h
			- c_{mh} m^2 h
			- \half k_{\mu\nu} \cF_{\mu\nu}  \ \\
	\lag_4^{tree} & = &  - {1 \over 4 \llambda}
				(\n_\mu \cU^T \n_\mu \cU)^2
			+ {1\over 2 \llambda} h (\n_\mu \cU^T \n_\mu \cU)
			- {1\over 4 \llambda} h^2
			- c_{hh} h^2  \\
	\lag_6^{tree} & = &  -  {1 \over 8 m^2 \llambda}
			( \n_\mu \cU^T \n_\mu \cU - h) \Box
			( \n_\nu \cU^T \n_\nu \cU - h) \ .
\end{eqnarray}

Now one can evaluate the one-loop contribution to the generating functional in
the effective theory, using the technique described in the previous section. At
order~$\llambda^0$, the matching relation~(\ref{match1}) is of the form
\begin{eqnarray} \label{match2}
\lefteqn{ \intdx \lag_\sigma   +   \half \ln\det d +
	      \half \ln\det \left( 1 - D^{-1} \delta^T d^{-1} \delta \right) }
				\nonumber \\
                    & & +    \half \ln\det D
                     +  \half \ln\det \left( d_M + \PT \sigma \PT
			 		\right)    \\
& = & \intdx (\lag_2 + \lag_4 + \lag_6)
                       +  \half \ln\det \bD
                       +  \half \ln\det \left( d_M + \PT \bar \sigma
					   \PT \right) \nonumber \ .
\end{eqnarray}
The quantities on the left hand side were defined in
Eqs.~(\ref{doperatorsfirst})-(\ref{doperatorslllast}), (\ref{d_M}) and
(\ref{sigma}), whereas those on the right hand side are given by
\bea
	\bD & = & -\Box
		+ \bX_{\mu\nu} \partial_\mu \partial_\nu
		+ (\partial_\mu \bX_{\mu\nu}) \partial_\nu \\
        \bar \sigma_{\mu\nu} & = & - M^2 \bar X_{\mu\nu}
                        - \bDelta^T_\mu \bD^{-1} \bDelta_\nu \\
	\bDelta_\mu   & = & - M \partial_\mu
			    + M \bX_{\mu\nu} \partial_\nu
		            + M (\partial_\nu \bX_{\nu\mu}) \\
	\bX_{\mu\nu}  & = & {1\over m^2} \left( 2 \cf_\mu \cf_\nu
			+ \delta_{\mu\nu}
			(\n_\rho \cU^T \n_\rho \cU - h) \right) \ .
\eea
Note that the operators on the right hand side of the matching
relation~(\ref{match2}) involve the solutions of the equations of motion of the
effective theory, while those on the left hand side depend on the solutions of
the equations of motion of the full theory. At the stationary point, however,
the corresponding corrections are of second order in the shift of the fields
and beyond the present accuracy. Thus, our notation will not distinguish
between the two solutions.

The last two terms on both sides of Eq.~(\ref{match2}) contain non-local
contributions corresponding to loops which involve only the light degrees of
freedom. They drop out of the matching relation. In the present case, however,
the situation is even simpler. There are no non-local contributions up to
order~$p^6$. This is a special property of the case $N=2$ of the gauged
linear $O(N)$ sigma model. At order~$p^2$ the effective Lagrangian describes a
free massive vector boson. Interaction only shows up at higher
order. Furthermore, there are no loop contributions in the effective theory at
order~$p^4$. Thus, in our case all low-energy constants at order~$p^2$
and~$p^4$ are finite.

The leading contributions of both~$\ln\det D$ and~$\ln\det\bar D$ are of
order~$p^8$. Thus, there is only one loop contribution in the effective theory
of the abelian Higgs model at order~$p^6$, the tadpole graph of the gauge
field:
\begin{equation} \label{tadpole}
\half \ln\det \left( 1 + d_M^{-1} \PT \bar \sigma \PT \right)  =
	- {M^2\over 2} G_{\mu\nu}(0) \intdx \bX_{\mu\nu}
			 + \order(p^8) \ .
\end{equation}
The free propagator of the vector field is of the form
\begin{equation}
	G_{\mu\nu}(x-y) \doteq
	\left<x\right| d_M^{-1} \PT_{\mu\nu} \left| y \right> \ .
\end{equation}

For completeness sake, we list all one-loop corrections to the generating
functional of the full theory which will contribute to the effective Lagrangian
up to the order $p^6$. First we define
\bea
	d	& = & d_m+ \sigma_m\\
	d_m 	& = & - \Box + 2m^2 \\
	D       & = & D_0 + \sigma_0 \\
	D_0	& = & - \Box \ .
\eea
One obtains the following terms which involve only the propagator of the
massive mode:
\be
	\half \ln\det d  = \half \ln\det d_m
		+ \half \Tr \left( d_m^{-1} \sigma_m \right)
		- \fourth \Tr \left( (d_m^{-1} \sigma_m)^2 \right) \nonumber\\
 		+ \sixth \Tr \left( (d_m^{-1} \sigma_m)^3 \right)
			\label{dloops} \ .
\ee
The second term, a tadpole graph, is of order $p^2$, whereas the
third and fourth traces are of order $p^4$ and $p^6$, respectively.

%\noindent
Mixed loops, which contain Higgs and Goldstone boson propagators, are given by:
\bea
	\lefteqn{ \hspace{-1.3cm} \half \ln\det \left( 1 - D^{-1} \delta^T
		d^{-1} \delta \right) =
		- \half \Tr \left( \delta D_0^{-1} \delta^T d_m^{-1} \right)
		+ \half \Tr \left( \delta D_0^{-1} \delta^T d_m^{-1} \sigma_m
				   d_m^{-1} \right) } \nonumber \\
& & 		- \half \Tr \left( \delta D_0^{-1} \delta^T
				   (d_m^{-1} \sigma_m)^2 d_m^{-1} \right)
		+ \half \Tr \left( \delta D_0^{-1} \sigma_0 D_0^{-1} \delta^T
				   d_m^{-1} \right) \nonumber \\
& &  		- \fourth \Tr \left( (\delta D_0^{-1} \delta^T d_m^{-1})^2
				   \right)
 		+ \half \Tr \left( (\delta D_0^{-1} \delta^T d_m^{-1})^2
				   \sigma_m d_m^{-1} \right) \nonumber \\
& & 		- \sixth \Tr \left( (\delta D_0^{-1} \delta^T d_m^{-1})^3
				   \right) \label{Ddloops}\ .
\eea
Here only the first term leads to contributions of order~$p^2$. The second
and fifth term are of order~$p^4$, while all other terms are of
order~$p^6$.

Finally, the following terms involve the gauge boson propagator:
\bea \label{gaugeloops}
	\lefteqn{ \half \ln\det \left( 1 + d_M^{-1} \PT \sigma \PT \right) =
 	 \half \Tr \left( d_M^{-1} \PT_{\mu\nu} M^2 (R^2 - 1) \right) }
	\nonumber \\
	& & -  \half \Tr \left( d_M^{-1} \PT_{\mu\nu} \delta^T_\nu d_m^{-1}
		\delta_\rho \right)
 	+ \half \Tr \left(
		d_M^{-1} \PT_{\mu\nu} \delta^T_\nu d_m^{-1} \sigma_m d_m^{-1}
		\delta_\rho \right) \nonumber \\
& & 	- \half \Tr \left(
		d_M^{-1} \PT_{\mu\nu} \delta^T_\nu d_m^{-1} \delta
		D_0^{-1} \delta^T d_m^{-1} \delta_\rho \right)  \ .
\eea
The second term, involving the mixed two-point function with a propagator of
the gauge field and the Higgs boson already contributes at order~$p^4$. The
tadpole graph and the three- and four-point functions contribute only at
order~$p^6$.

Techniques to evaluate the low-energy expansion of the traces in
Eqs. (\ref{dloops}), (\ref{Ddloops}), and~(\ref{gaugeloops}) are discussed in
Ref.~\cite{LSM}. We obtain the following result for the effective Lagrangian of
the abelian Higgs model up to order~$p^6$ and~$\llambda^0$:
\bea
\lag_2 & = & 	\left( {1 \over 4 \llambda}
		- 3 \pole + {1\over 16 \pi^2} \fourth \right)
		2m^2 \n_\mu \cU^T \n_\mu \cU
		+ {1 \over 4 \g^2} \cF_{\mu\nu} \cF_{\mu\nu}
		\label{lag2_bare} \\
\lag_4 & = & 	\left( 3 \pole + {1\over 16 \pi^2} \fourth  \right)
		M^2 \n_\mu \cU^T \n_\mu \cU \nonumber \\
& & 		- \left( {1\over 6} \pole + {1\over 16\pi^2} {1\over 72}
		\right)
		\cF_{\mu\nu} \cF_{\mu\nu}
		\nonumber \\
& & 		- \left({1 \over 4 \llambda}
		- 5 \pole - {1\over 16\pi^2} {19 \over 12} \right)
		(\n_\mu \cU^T \n_\mu \cU)^2
		\label{lag4_bare} \\
\lag_6 & = & 	\left( 3 \pole + {1\over 16 \pi^2} \fourth \right)
		{M^4 \over 2m^2} \n_\mu \cU^T \n_\mu \cU \nonumber \\
& & 		+ \left( {1 \over 3} {1\over 16\pi^2} \right)
		{M^2 \over 2m^2} \cF_{\mu\nu} \cF_{\mu\nu} \nonumber \\
& & 		- \left( 6 \pole + {1\over 16\pi^2} {9 \over 2} \right)
		{M^2 \over 2m^2} (\n_\mu \cU^T \n_\mu \cU)^2
		\nonumber \\
& & 		- {1\over 16\pi^2} {233\over 72}
		{1\over 2m^2} (\n_\mu \cU^T \n_\mu \cU^T)^3 \nonumber \\
& & 		- \left( {1 \over 2 \llambda} - 14 \pole - {1\over 16\pi^2}
		{50 \over 9} \right)
		{1\over 2m^2} (\n_\mu \cU^T \n_\mu \cU)
		(\n_\nu \cU^T \n_\rho \n_\rho \n_\nu \cU) \nonumber \\
& & 		- \left( {1 \over 2 \llambda} - 14 \pole - {1\over 16\pi^2}
		{58 \over 9} \right)
		{1\over 2m^2} (\n_\mu \cU^T \n_\mu \cU)
		(\n_\nu \n_\rho \cU^T \n_\nu \n_\rho \cU) \nonumber \\
& & 		- {1\over 16\pi^2} {25\over 36}
		{1\over 2m^2} (\n_\mu \cU^T \n_\mu \cU)
		(\n_\nu \n_\rho \cU^T \n_\rho \n_\nu \cU) \nonumber \\
& & 		- {1\over 16\pi^2} {1\over 24}
		{1\over 2m^2}
		\n_\mu \n_\mu \n_\rho \cU^T \n_\nu \n_\nu \n_\rho \cU
		\nonumber \\
& & 		+  {1\over 16\pi^2} {1\over 36}
		{1\over 2m^2} (\n_\mu \n_\mu \n_\rho \cU^T \cU)
		(\cU^T \n_\nu \n_\nu \n_\rho \cU) \ ,
		\label{lag6_bare}
\eea
where
\be
	\pole  \doteq  {1\over 16\pi^2} \mu^{d-4} \left( {1\over d-4} -
		\half \left( \ln 4 \pi + \Gamma^{\prime}(1) + 1\right) \right)
		+ {1\over 32 \pi^2} \ln \left( {2m^2 \over \mu^2} \right) \ .
\ee
Note that we have switched off the external sources $h$ and $k_{\mu\nu}$. An
effective field theory analysis to one loop, based on this Lagrangian, will
yield the low energy expansion of the one-loop approximation in the abelian
Higgs model up to order~$p^6$. Two-loop corrections in the effective theory
will also be of order~$p^6$. However, they are of higher order in~$\llambda$.

\rsection{On the Loop Expansion}

In this section we will briefly discuss loop contributions of arbitrary order.
To go beyond the one-loop approximation, one has to include higher orders in
the fluctuation~$y$ around the classical solution~$(\cphi^A, \cA_\mu)$ in
Eq.~(\ref{W_pathint}). One obtains
\begin{equation}  \label{fullexpansion}
	\intdx \lag_\sigma(\phi,A) = \intdx \left(
		\lag_\sigma(\cphi,\cA_\mu)
		+ \half y^T \tildeD y + \lag_\sigma^{[3]}
				+ \lag_\sigma^{[4]} \right) \ ,
\end{equation}
where $y^T = ( f, q_\mu )$, and
\begin{eqnarray} \label{firstvertex3}
 \lag_\sigma^{[3]} & = &
	\g q_\mu (\cnabla_\mu f^T) T_e f
	+ \g^2 q_\mu q_\mu \cphi^T f
	+ \llambda f^T f \cphi^T f \\
 \lag_\sigma^{[4]} & = &
	{ \g^2\over2 }  q_\mu q_\mu f^T f + {\llambda\over4} (f^T f)^2 \ .
	\label{firstvertex4}
\end{eqnarray}
The perturbative expansion of the generating functional to arbitrary order is
then given by
\bea \label{genfuncgen}
 e^{-W_\sigma[h,k_{\mu\nu}]} & = &
	e^{-\intdx\lag_\sigma(\cphi,\cA_\mu)}
		\int\d\mu[\phi,A] e^{- \half\intdx y^T \tildeD y}
	\Biggl( 1 - \intdx \lag_\sigma^{[4]}
	\nonumber \\
& & 	\hspace*{4cm} + \left. \half \left( \intdx \lag_\sigma^{[3]} \right)^2
	+ \ldots \right) \ .
\eea
Note that any 3- (4-)vertex can emit up to three (four) zero-modes. To get
a first idea about how diagrams with zero-modes have to be treated, let us
consider diagrams where one 3-vertex emits exactly three zero-modes. To make
things well-defined, one may add an infinitesimal mass term $y^T\epsilon y$ to
the quadratic form in the exponent of Eq.~(\ref{genfuncgen}).
Substituting
\begin{equation}
	y = \sum_{n} b_n P \alpha_n
\end{equation}
for every fluctuation, one obtains
\begin{equation} \label{emitthree}
 \intdx \lag_\sigma^{[3]} = {g^3 \over 3} \sum_{mno} b_m b_n b_o \intdx
	(\partial_\mu J_\mu) \alpha_m \alpha_n \alpha_o \ .
\end{equation}
This expression vanishes identically due to current
conservation~(\ref{eqmo2cc}). In order to derive Eq.~(\ref{emitthree}), it
was necessary to sum over all zero-modes. Individual diagrams with
zero-modes may well yield non-zero contributions to the generating
functional. However, the example shows that all diagrams where one particular
3-vertex emits three zero-modes cancel each other. The situation is more
complicated if 3-vertices emit less than three zero-modes or if 4-vertices
are involved. In such cases, a cancellation can only be expected between all
diagrams involving the same number of zero-modes.

This complication can be traced back to our parametrization~(\ref{fluct1phi})
and (\ref{fluct1A}) of the quantum fluctuations. The Lagrangian is invariant
under gauge transformations of the form~(\ref{gaugetrafo1})
and~(\ref{gaugetrafo2}), acting on the fields~$(\phi, A_\mu)$. For fixed
classical fields~$(\cphi, \cA_\mu)$, they imply the following transformations
of the quantum fluctuations:
\begin{eqnarray}
	f^{\prime} 	& = & e^{\omega T_e} f
			 + \left( e^{\omega T_e} - 1 \right) \cphi \\
	q_\mu^{\prime} 	& = & q_\mu - {1\over \g} \partial_\mu \omega \ .
\end{eqnarray}
Only infinitesimal gauge transformations correspond to zero-modes of the
operator~$\tildeD$:
\begin{equation}
	y^{\prime} = y + {1\over g} P \omega + \order({\omega^2}) \ ,
\end{equation}
where~$y$ itself is also treated as order~$\omega$. Thus, the quadratic form on
the right hand side of Eq.~(\ref{fullexpansion}) is gauge invariant only at
leading order in~$\omega$. Under finite gauge transformations, only the sum of
the last three terms is invariant, which leads to nontrivial identities between
the vertices.

In order to show that the zero-mode contributions cancel each other, it is
suitable to introduce another parametrization for the quantum fluctuations:
\begin{eqnarray}
	\phi  & = & (1 + \nf_1) e^{ \nf_2 T_e} \cphi \  \label{newparam1} \\
	A_\mu & = & \cA_\mu + g q_\mu \ . \label{newparam2}
\end{eqnarray}
With
\begin{equation}
	\nf_i = {\sqrt{\llambda}\over m} { \nft_i \over R }
\end{equation}
one obtains
\begin{equation}  \label{sfullexpansion}
	\intdx \lag_\sigma(\phi,A) = \intdx \left(
		\lag_\sigma(\cphi, \cA_\mu)
		+ \half \ty^T \ttD \ty
		+ \tlag_\sigma^{[3]}
		+ \tlag_\sigma^{[4]} \right) \ ,
\end{equation}
where $\ty^T = ( \nft_1 , \nft_2, q_\mu )$,
\begin{equation}
	\ttD =  \left( \begin{array}{ccc}
                d & \delta & \delta_\nu \\
		\delta^T & D & \Delta_\nu \\
		\delta_\mu^T &  \Delta_\mu^T & D_{\mu\nu}^{qq}
\end{array}\right)  \ ,
\end{equation}
and
\begin{eqnarray} \label{secondvertex3}
 \tlag_\sigma^{[3]} & = &
	\sqrt{\llambda} m R \nft_1^3
	+ {\sqrt{\llambda} \over m} M^2 R \nft_1 ( q_\mu + \partial_\mu \left(
						{\nft_2\over M R} \right) )^2
		\nonumber \\
      & & \qquad \mbox{} - {\sqrt{\llambda} \over m} M \nft_1^2 f_\mu ( q_\mu
		+  \partial_\mu \left( {\nft_2\over M R} \right) ) \ ,\\
 \tlag_\sigma^{[4]} & = &
	\fourth \llambda \nft_1^4
        + \half {\llambda\over m^2} M^2 \nft_1^2 ( q_\mu + \partial_\mu \left(
						{\nft_2\over M R} \right) )^2 \
. \label{secondvertex4}
\end{eqnarray}
The components of the operator~$\ttD$ are given in
Eqs.~(\ref{doperatorsfirst})-(\ref{doperatorsllast}). In the new
parametrization zero-modes are of the form
\begin{equation}
      \ty^T = ( 0, M R \omega, -\partial_\mu \omega ) \ ,
\end{equation}
and thus correspond to gauge transformations of arbitrary size. Furthermore,
the combination
\begin{equation}
	q_\mu + \partial_\mu \left( {\nft_2\over M R} \right)
\end{equation}
is gauge invariant, and vanishes identically if a zero-mode is inserted. Thus,
in the new parametrization it is obvious that diagrams with zero-modes do not
contribute to the generating functional. The vertices in
Eqs.~(\ref{secondvertex3})-(\ref{secondvertex4}) are already grouped
appropriately. If the operator~$\ttD$ is diagonalized by the
transformation~(\ref{diagtrafo}), applied to the fields~$(\nft_1, \nft_2,
q_\mu)$, the vertices in~$\tlag_\sigma^{[3]}$
and~$\tlag_\sigma^{[4]}$ change according to
\begin{eqnarray}
	\nft_1(x) & \rightarrow & \intdy \left<x\right| \left( \nft_1
		+ v^{1,2} \nft_2 + v^{1,3}_\mu q_\mu \right)\left|y\right> \ ,
	\\
 M ( q_\mu + \partial_\mu \left( {\nft_2\over M R} \right) )(x)
 & \rightarrow & \intdy \left<x\right| \left( v^{2,2}_\mu \nft_2
		+ v^{2,3}_{\mu\nu} q_\nu \right)\left|y\right> \ ,
\end{eqnarray}
with
\begin{eqnarray}
	v^{1,2} & = & -d^{-1} \delta \\
	v^{1,3}_\mu & = & -d^{-1} \delta_\mu + d^{-1}\delta\Theta^{-1}\X_\mu\\
	v^{2,2}_\mu & = & {1\over M R^2} \Delta^T_\mu \\
	v^{2,3}_{\mu\nu} & = & M \delta_{\mu\nu}
		- {1\over M R^2} \Delta_\mu^T\Theta^{-1}\X_\nu \ .
\end{eqnarray}
The zero-modes are now of the form~(\ref{simplezeromode}) and correspond to
longitudinal fluctuations of the gauge field. They are not emitted from any
vertex, since
\begin{equation}
	v^{1,3}_\mu \partial_\mu \alpha_m =
	v^{2,3}_{\mu\nu} \partial_\nu \alpha_m  = 0 \ .
\end{equation}
The full gauge field propagator entering loop graphs is given by the inverse
of the operator~$\varD_{\mu\nu}$, restricted to the subspace of nonzero-modes.
Thus, the calculation of loop contributions of arbitrary order is boiled down
to the evaluation of Gaussian integrals. One obtains, for example,
\begin{equation}
	\int \d\mu[q] q_\mu^T(x) q_\nu^T(y) e^{- \half\intdx q \varD q }
		= \langle x| \varD^{-1\prime}_{\mu\nu} | y \rangle
		\sqrt{\det( \hat P^T \hat P)}/\sqrt{\det^\prime(\varD)} \ ,
\end{equation}
where~$q_\mu^T$ is the transverse component of the fluctuation~$q_\mu$,
i.e.,~$q^T = \PT q$, and
\begin{equation}
	\varD \varD^{-1\prime} = \varD^{-1\prime} \varD = \PT \ .
\end{equation}

We do not need to go any further. The differential operator $\ttD$ as well as
the vertices in the Lagrangians~$\tlag_\sigma^{[3]}$ and~$\tlag_\sigma^{[4]}$
depend on the coupling constant~$\g$ only through the gauge boson
mass~$M$. Furthermore, $\tlag_\sigma^{[3]}$ and~$\tlag_\sigma^{[4]}$ are
proportional to~$\sqrt{\llambda}$ and~$\llambda$, respectively. Hence, $n$-loop
contributions count as order~$\llambda^{n-1}$. This includes tree-level and
one-loop contributions as discussed in the previous section. Tree-level
corresponds to $n=0$. The low energy expansion at a given loop level generates
only powers of the momentum and the gauge boson mass~$M$. Hence, it does not
mix up the counting of~$\llambda$. The order of $n$-loop contributions can
already be inferred from Eqs.~(\ref{firstvertex3})
and~(\ref{firstvertex4}). The discussion in this Section shows that the
proper treatment of the zero-modes does not change this result.

\rsection{Renormalization}

To render the generating functional~(\ref{1loopgf}) of the abelian Higgs model
finite, one has to renormalize the bare constants $m, \llambda, \g, c_{mh},
c_{hh}$, the scalar field~$\phi$, and the source $h$ before the regulator can
be removed. There is no wave function renormalization of the vector field on
account of gauge invariance (cf.~Eq.~(\ref{cov_deriv})). The ultraviolet
divergences are related to the poles of the d-dimensional determinants which
appear in the generating functional for $d=0,2,4,\dots$ .  For a differential
operator $D$ of the form
\begin{equation} \label{Dform}
	D = -D_\mu D_\mu + \sigma \ , \qquad
		D_\mu = \partial_\mu  + \Gamma_\mu \ ,
\end{equation}
the pole term of the determinant at $d=4$ is given by
\begin{equation}
	\half \ln\det D = {1\over d-4} {1\over 16\pi^2}
	\intdx tr \left( {1\over12} \Gamma_{\mu\nu} \Gamma_{\mu\nu}
			+ \half \sigma^2 \right) + \order(1) \ ,
\end{equation}
with
\begin{equation}
	\Gamma_{\mu\nu} = \left [ D_\mu , D_\nu \right ] \ .
\end{equation}
This identity can readily be derived~\cite{LSM} using the heat kernel
method~\cite{heatkernel}.  Since the operators $P^T P$ and $\tildeD + P
P^T$ are both of the form~(\ref{Dform}), it is straightforward to verify that
the poles in the generating functional~(\ref{1loopgf}) are removed by the
following renormalization prescriptions:
\begin{eqnarray}
	Z_\phi & = & 1 - 8 \g_r^2 \left( \pole + \delta z \right)
	\label{Z_phi1} \\
	m^2 & = & m_r^2 \left( 1 - 2 \left( 4 \llambda_r + \g_r^2\right)
		\left( \pole + \delta m^2\right) - ( Z_\phi-1 ) \right)
	\label{m_r} \\
	\llambda & = & \llambda_r \left( 1 - \left( 20 \llambda_r + 4 \g_r^2
			+ 6 {\g_r^4\over\llambda_r}  \right)
				\left( \pole + \delta \llambda \right)
			- 2 (Z_\phi - 1)  \right) \label{ll_r} \\
	\g^2 & = & \g_r^2 \left( 1 - {2\over3} \g_r^2
			\left( \pole + \delta \g^2\right) \right)
			\label{g_r} \\
 	h & = & c_r  h_r \\
	c_{hh} & = & c_{hh,r}  + \left( \pole + \delta c_{hh} \right)
		- 2 c_{hh,r} (c_r - 1) \\
	c_{mh} & = & c_{mh,r}  + 2 \left( \pole + \delta c_{mh} \right)
		- c_{mh,r} (c_r - 1)
		- c_{mh,r} \left( { m^2 - m_r^2 \over m_r^2 } \right)
		\nonumber \\
& &
\end{eqnarray}
with
\begin{equation}
	c_r = 1 - 2 \left( 4 \llambda_r + \g_r^2 \right) \left( \pole
		+ \delta c_r \right) - ( Z_\phi - 1) \ ,
\end{equation}
and
\begin{equation}
	\phi = Z_\phi^{1/2} \phi_r \label{Z_phi2} \ .
\end{equation}
Eqs.~(\ref{Z_phi1})-(\ref{Z_phi2}) introduce the finite but otherwise
completely arbitrary constants~$\llambda_r,$ $\g_r, \, m_r, \, Z_\phi, \,
h_r, \, c_{hh,r}, \, c_{mh,r}, \, \delta \llambda, \, \delta \g^2, \, \delta
m^2, \, \delta z, \, \delta c_{hh}, \, \delta c_{mh},$ and~$\delta c_r$. They
are determined by the renormalization scheme. In the following we will express
the effective Lagrangian~(\ref{lag2_bare})-(\ref{lag6_bare}) in terms of the
physical masses~$M_H$ and~$\MZ$ of the Higgs and gauge boson, as well as the
parameter~$g_{res}$ defined by
\begin{equation}  \label{gresdef}
	\langle 0 | F_{\mu\nu}(0) | k, \epsilon^{(\sigma)} \rangle
		\doteq i g_{res} \left( k_\mu \epsilon_\nu^{(\sigma)}
				- k_\nu \epsilon_\mu^{(\sigma)} \right) \ .
\end{equation}
Here,~$| k, \epsilon^{(\sigma)}\rangle$ denotes a physical state of the massive
gauge boson with momentum~$k$ and polarization~$\epsilon^{(\sigma)}$.
The physical masses are determined by the pole positions of the two-point
functions
$ \langle 0 | T (\phi^T \phi)(x) (\phi^T \phi)(y) | 0 \rangle $
and
$ \langle 0 | T F_{\mu\nu}(x) F_{\rho\sigma}(y)| 0 \rangle . $
Furthermore, according to Eq.~(\ref{gresdef}) the quantity~$g_{res}^2$ is
determined by the residue of the two-point function of the field strength.

All of these statements can conveniently be summarized at tree level by the
following result for the generating functional $W_\sigma[h,k_{\mu\nu}]$,
expanded up to second order in the sources:
\bea
W_\sigma[h,k_{\mu\nu}]^{tree} & = &
			- \left( {m^2 \over 2 \llambda} + c_{mh} m^2 \right)
			\intdx \ h  - c_{hh}  \intdx \ h^2
			\nonumber \\
& & 			- \left( {m^2 \over 2 \llambda} \right)
			\intdx \d^d y \ h(x)
			\left<x\right| d_m^{-1} \left| y \right> h(y) \\
& &
			- \half \g^2 \intdx \d^d y \
				(\partial_\mu k_{\mu\nu})(x)
				\left<x\right| d_M^{-1} \left| y \right>
			  	(\partial_\rho k_{\rho\nu})(y) \ .
			\nonumber
\eea
If the one-loop corrections in the abelian Higgs model are taken into account,
one obtains
\bea
        M_H^2
& = & 2m^2 \left\{ 1 + \llambda \left[
		8 \left( 1 - {3 \over 2} {M^2 \over 2m^2} \right) \pole
		\right. \right. \nonumber \\
& & 		\left. \left. + {1\over 16\pi^2} \left( - 10 +
			3 \sqrt{3} \pi  + 2 {M^2 \over 2m^2}
			\right) + \order(p^4)
		\right] \right\} \label{M_Higgs} \\
        \MZ^2      		& = & M^2 \left\{ 1 + \llambda \left[
		- 12 \left( 1 - {10\over 9} {M^2 \over 2m^2}
			+ 2 {M^4 \over (2m^2)^2} \right) \pole \right. \right.
		\nonumber \\
& & 		\left. \left.  + {1\over 16 \pi^2} \left( 1 +
			{10 \over 9}  {M^2 \over 2m^2}
			- \left( {65\over 6} + 18 \,
			\ln \left( {M^2 \over 2m^2} \right) \right)
			{M^4 \over (2m^2)^2} \right)
			+\order(p^6) \right]  \right\}
		\nonumber \\
& & 		\label{M_Gaugeboson} \\
        \g_{res}^2 		& = & g^2 \left\{ 1 + \g^2 \left[ {2 \over 3}
		\pole +  {1 \over 16\pi^2} \left( {1\over 18}
		- {3\over 2} {M^2 \over 2m^2}\right)
		+\order(p^4)\right]  \right\}
		\ . \label{g_res}
\eea
These expressions will be finite, if the renormalization
prescriptions~(\ref{Z_phi1})-(\ref{g_r}) are inserted on the right hand side.
One obtains the same result for $\MZ$ and $\g_{res}$ in the
effective field theory if all terms up to order $p^6$ in $\lag_{eff}$ are
taken into account.

In terms of these physical parameters, the effective Lagrangian reads
\bea \label{lageff_ren}
\lag_{2} 	& = &
		\half {\MZ^2 \over \g_{res}^2}
		\n_\mu \cU^T \n_\mu \cU
		+ {1 \over 4 \g_{res}^2} \cF_{\mu\nu} \cF_{\mu\nu}
		\\
\lag_4 & = & 	- \left( { \MZ^2 \over 2 M_H^2 \, \g_{res}^2}
		- {52 - 9 \sqrt{3} \pi\over 192\pi^2} \right)
		(\n_\mu \cU^T \n_\mu \cU)^2 \\
\lag_6 & = & 	\left( 9 \poleMH + {1\over 16 \pi^2} \left[ {53\over 24} +
		{9\over 2} \ln \left( {\MZ^2 \over M_H^2} \right) \right]
		\right)
		{\MZ^4 \over M_H^2} \n_\mu \cU^T \n_\mu \cU
		\nonumber \\
& & 		- {1\over 16\pi^2} {1\over 24}
		{\MZ^2 \over M_H^2} \cF_{\mu\nu} \cF_{\mu\nu}
  		- {1\over 16\pi^2} {19 \over 4}
		{\MZ^2 \over M_H^2} (\n_\mu \cU^T \n_\mu \cU)^2
		\nonumber \\
& & 		- {1\over 16\pi^2} {233\over 72}
		{1\over M_H^2} (\n_\mu \cU^T \n_\mu \cU^T)^3 \nonumber \\
& & 		- \left( {\MZ^2 \over M_H^2 \, g_{res}^2} -
		{289 - 54 \sqrt{3} \pi \over 288\pi^2} \right)
		{1\over M_H^2} (\n_\mu \cU^T \n_\mu \cU)
		(\n_\nu \cU^T \n_\rho \n_\rho \n_\nu \cU) \nonumber \\
& & 		- \left( {\MZ^2 \over M_H^2 \, \g_{res}^2} -
		{305 - 54 \sqrt{3} \pi \over 288\pi^2} \right)
		{1\over M_H^2} (\n_\mu \cU^T \n_\mu \cU)
		(\n_\nu \n_\rho \cU^T \n_\nu \n_\rho \cU) \nonumber \\
& & 		- {1\over 16\pi^2} {25\over 36}
		{1\over M_H^2} (\n_\mu \cU^T \n_\mu \cU)
		(\n_\nu \n_\rho \cU^T \n_\rho \n_\nu \cU) \nonumber \\
& & 		- {1\over 16\pi^2} {1\over 24}
		{1\over M_H^2}
		\n_\mu \n_\mu \n_\rho \cU^T \n_\nu \n_\nu \n_\rho \cU
		\nonumber \\
& & 		+  {1\over 16\pi^2} {1\over 36}
		{1\over M_H^2} (\n_\mu \n_\mu \n_\rho \cU^T \cU)
		(\cU^T \n_\nu \n_\nu \n_\rho \cU) \ .
\eea
Note, that some terms of order $p^4$ have disappeared from the effective
Lagrangian. This is due to the fact that the relations (\ref{M_Higgs})  -
(\ref{g_res}) between bare and physical parameters are not homogeneous in the
order of the momentum.  Thus, renormalizing the effective Lagrangian mixes
various low energy constants.

The contribution from the tadpole graph~(\ref{tadpole}) can be accounted for by
subtracting the term
\begin{equation}
	       9\left(\poleMH + {1\over 16 \pi^2} \left[ {1\over 4} +
		{1\over 2} \ln \left( {\MZ^2 \over M_H^2} \right) \right]
		\right)
		{\MZ^4 \over M_H^2} \n_\mu \cU^T \n_\mu \cU
\end{equation}
from the effective Lagrangian given above. This yields a finite result, since
there are no other one-loop contributions in the effective theory of the
abelian Higgs model up to order~$p^6$.

The representation of the low energy constants in terms of the
quantities~$M_H, \MZ$, and~$g_{res}$ can be used to estimate the value of the
Higgs boson mass where strong coupling sets in. Using Standard Model values
for the gauge boson mass~$\MZ$ and the coupling~$g_{res}$, one obtains the
value~$M_H\sim 4\mbox{\rm TeV}$. At this point one-loop corrections to the low
energy constants are of the same size as the tree level contributions.  This
number is larger than the corresponding estimate in the non-abelian case,
since the low energy constants depend on the rank of the group~$O(N)$. For
larger rank this estimate is smaller.

Finally, we note that in the limit $\g_{res}^2 \to 0$ and $\MZ^2 \to 0$, such
that
\be
        {\MZ^2 \over \g_{res}^2} \doteq F^2 = \mbox{const} \ ,
\ee
the low energy constants for the ungauged $O(2)$ linear sigma model~\cite{LSM}
are recovered.

\rsection{Summary and Discussion}

In this work we have discussed a manifestly gauge invariant\footnote{The
literature on the background field effective action distinguishes between gauge
invariance and gauge independence. In this work the phrase gauge invariant
includes both meanings.} approach to analyze the low energy structure of the
gauged linear sigma model. In the spontaneously broken phase the spectrum of
this model contains one massive scalar particle, the Higgs boson. If its mass
is large enough, the low energy structure of the linear sigma model can
adequately be described by an effective field theory. Furthermore, if the
couplings are not too strong, perturbative methods apply. Thus, there exists an
intermediate range for the mass of the Higgs boson where both the effective
field theory description and the perturbative treatment in the coupling
constants are valid.  Our analysis was concerned with this case. In particular,
we discussed a functional technique to evaluate the gauge invariant effective
Lagrangian which describes the effective field theory of the linear sigma
model. The advantage of our approach is that it explicitly reflects the
symmetry properties of the underlying theory, i.e., it is manifestly gauge
invariant.

In order to avoid technical difficulties we have chosen the abelian Higgs model
as a simple example to demonstrate our method. It corresponds to the case
$N=2$ of the gauged linear~$O(N)$ sigma model. However, our analysis does not
rely on any particular property of the abelian theory. We will see below that
it can readily be extended to the non-abelian case.

The effective Lagrangian of the linear sigma model is a sum of gauge invariant
terms with an increasing number of covariant derivatives and gauge boson mass
factors, corresponding to an expansion in powers of the momentum and the mass.
Note that the covariant derivative, the gauge boson mass~$M$, and the gauge
coupling~$\g$ all count as quantities of order~$p$. Thus, the low energy
expansion is carried out at a fixed ratio~$p^2/M^2$, which ensures that all
light-particle singularities are correctly reproduced. Furthermore, if the
coupling~$\llambda$ of the scalar field is small enough, the low energy
constants in the effective Lagrangian admit an expansion in powers of this
quantity, corresponding to the loop expansion in the full theory. $n$-loop
Feynman diagrams in the abelian Higgs model yield corrections of
order~$\llambda^{n-1}$ to the low energy constants. Hence, in the intermediate
range of the Higgs boson mass the accuracy of the effective field theory
description is controlled by the order of both the momentum and the coupling
constant~$\llambda$. Note, however, that the correspondence between powers of
the coupling~$\llambda$ and the order of loops in the full theory becomes
meaningless in the strong coupling limit; i.e., for very large values of the
Higgs mass. In this case, the functional relationships between the low energy
constants and the coupling of the scalar field cannot be evaluated with
perturbative methods.

In the intermediate range for the Higgs boson mass one can make use of this
correspondence and separately specify the order of the momentum and the scalar
coupling to be taken into account, depending on the energy scale and the value
of the Higgs mass relevant for a given analysis. This may turn out to be
useful in the effective field theory analysis of electroweak symmetry
breaking, where the Standard Model with one scalar doublet serves as a
reference point.  For values of the Higgs mass close to the strong coupling
region, one may need to go beyond the next-to-leading order approximation for
the low energy constants. In this work we have evaluated the effective
Lagrangian of the abelian Higgs model up to order~$p^6$ and the low energy
constants up to order~$\llambda^0$. Thus, an effective field theory analysis
based on this Lagrangian yields the low energy expansion of the corresponding
one-loop approximation in the full theory up to order~$p^6$. Finally, we have
expressed the effective Lagrangian in terms of the physical masses of the
Higgs and the gauge boson, and the gauge coupling. From this representation of
the low energy constants, one can estimate the intermediate range for the mass
of the Higgs boson. Using Standard Model values, this range turns out to be~$
0.1 \mbox{\rm TeV}\lsim M_H \lsim 4 \mbox{\rm TeV}$. For larger values of this
mass, one enters the strong coupling regime, and perturbation theory with
respect to the parameter~$\llambda$ is no longer possible. For smaller values
of the Higgs boson mass, the effective field theory description ceases to be
valid.

It is crucial for a manifestly gauge invariant approach that the external
sources respect the symmetry properties of the theory. Thus, our analysis was
only concerned with Green's functions of gauge invariant operators. In the
abelian Higgs model this causes no restriction. In the spontaneously broken
phase, Green's functions of the scalar density~$\phi^T\phi$ and the field
strength~$F_{\mu\nu}$ have poles at the masses of the Higgs and gauge
boson. Thus, one can extract all $S$-matrix elements of the theory. One may
even couple an external vector field to the current~$(\nabla_\mu\phi)^T
T_e\phi$ and consider Green's functions involving this operator as well. For
large values of the Higgs mass, the low energy structure of these functions can
adequately be described by an effective field theory with a gauge invariant
effective Lagrangian. Thus, as far as the abelian Higgs model is concerned,
things work as smoothly as in the ungauged case.

In the non-abelian theory the situation is a little bit different. The field
strength and the currents are not gauge invariant. One cannot couple external
sources to these operators without breaking the symmetry. Thus, in order to
evaluate $S$-matrix elements in a gauge invariant framework one has to find
other sources, which may either create single particles or pairs. The latter
might be more appropriate for charged particles.  A particular kind of gauge
invariant source term which may help in this case is discussed in the context
of the Vilkovisky-DeWitt effective action~\cite{VDEA}.

The situation simplifies considerably if one is interested in the effective
Lagrangian. The only reason for us to couple an external source to the field
strength was to ensure that we manipulate non-vanishing classical fields during
intermediate steps of our calculation. However, this can also be achieved
without an external source simply by imposing non-trivial boundary conditions
on the field strength. In fact, the whole analysis in this article involved
only the solution of the equation of motion for the Higgs field. Thus, our
technique of evaluating the gauge invariant effective Lagrangian can be applied
to the non-abelian case without modification.

To determine $S$-matrix elements in the effective field theory, one may couple
source terms to this Lagrangian and consider Green's functions of the
corresponding operators. Since our renormalization of the low energy constants
in the non-abelian case will not know about these source terms, Green's
functions will generally be well defined only in the regularized theory. Thus,
one has to extract $S$-matrix elements before the regulator can be removed.
Once the effective Lagrangian is given, one may even use the Faddeev-Popov
ansatz to evaluate scattering amplitudes.

\section*{Acknowledgements}

We are grateful to H.~Leutwyler for many enlightening discussions and a
critical reading of this manuscript.  A.N.\ is furthermore indebted to
J.~Gasser and E.~Kraus for useful discussions.  A.S.\ would like to thank
H.~Georgi for clarifying communications. He is also grateful to the members of
the Harvard Physics Department for their kind hospitality.


\newpage
\begin{thebibliography}{44}


\bibitem{L_eff_SM}

	For reviews on the use of effective Lagrangians in electroweak
	physics, see

\artref
	{F.\ Feruglio}{Int.\ J.\ Mod.\ Phys.}{A8}{(1993) 4937;}
%	[hep-ph/9301281];

\artref
	{J.\ Wudka}{Int.\ J.\ Mod.\ Phys.}{A9}{(1994) 2301.}
%	[hep-ph/9406205].


\bibitem{heavy_Higgs}

\artref
	{T.\ Appelquist and C.\ Bernard}{Phys.\ Rev.}{D22}{(1980) 200;}

\artref
	{A.\ C.\ Longhitano}{Phys.\ Rev.}{D22}{(1980) 1166;}

\bookref
	{T.\ Appelquist, {\em in} {\rm ``Gauge Theories and Experiments at High
	Energies'' (K.\ C.\ Bowler and D.\ G.\ Sutherland, Eds.)}}{Scottish
	University Summer School in Physics, St.\ Andrews, 1980;}

\artref
	{A.\ C.\ Longhitano}{Nucl.\ Phys.}{B188}{(1981) 118.}


\bibitem{technicolor}

\artref
	{S.\ Weinberg}{Phys.\ Rev.}{D19}{(1979) 1277;}

\artref
	{L.\ Susskind}{Phys.\ Rev.}{D20}{(1979) 2619;}

\artref
       	{E.\ Farhi and L.\ Susskind}{Phys.\ Rep.}{74}{(1981) 277.}


\bibitem{LSM}
\artref
        {A.\ Nyf\/feler and A.\ Schenk}{{\rm
	%``Effective Field Theory of the Linear $O(N)$ Sigma Model'',
	Bern University,}}{BUTP-94/12,}{June 1994,
	{\it Ann. Phys. {\bf 241} {\rm (1995)}, in press,}} [hep-ph/9409436].


\bibitem{H_M}
\artref
	{M.\ J.\ Herrero and E.\ R.\ Morales}{Nucl.\ Phys.}{B418}{(1994) 431;}
%	[hep-ph/9308276];

\artref
	{M.\ J.\ Herrero and E.\ R.\ Morales}{Nucl.\ Phys.}{B437}{(1995) 319.}
%	[hep-ph/9411207].


\bibitem{E_M}
\artref
	{D.\ Espriu and J.\ Matias}{Phys.\ Lett.}{B341}{(1995) 332.}
%	[hep-ph/9407292].


\bibitem{D_GK}
\artref
	{S.\ Dittmaier and C.\ Grosse-Knetter}{\rm
	% ``Deriving Non-decoupling Effects of Heavy Fields from the Path
	% Integral: a Heavy Higgs Field in an $SU(2)$ Gauge Theory'',
	Bielefeld University,}{BI-TP-95-01,}{ Jan 1995} [hep-ph/9501285];

\artref
	{S.\ Dittmaier and C.\ Grosse-Knetter}{\rm
	% ``Integrating out the Standard Higgs Field in the Path Integral'',
	Bielefeld University,}{BI-TP-95-10,}{ May 1995} [hep-ph/9505266].


\bibitem{Polyakov}
\artref
	{A.\ M.\ Polyakov}{Nucl.\ Phys.}{B120}{(1977) 429.}


\bibitem{Corrigan_Goddard}
\bookref
	{E.\ Corrigan and P.\ Goddard, {\em in} {\rm ``Geometrical and
	Topological Methods in Gauge Theories'' (J.\ P.\ Harnad and S.\
	Shnider, Eds.)}}{Proceedings, Montreal, 1979; Lecture Notes in Physics
	129, Springer-Verlag, Berlin, 1980.}


\bibitem{ChPT}
\artref
        {S.\ Weinberg}{Physica}{A 96}{(1979) 327;}

\artref
	{J.\ Gasser and H.\ Leutwyler}{Ann.\ Phys.}{158}{(1984) 142.}


\bibitem{heatkernel}
\artref
	{J.\ Schwinger}{Phys.\ Rev.}{82}{(1951) 664;}

\bookref
	{P.\ B.\ Gilkey, ``The Index Theorem and the Heat Equation``}{Publish
	or Perish, Boston, 1974.}


\bibitem{VDEA}

	Reviews on the Vilkovisky-DeWitt effective action can be found in

\bookref
	{{\rm ``$TeV$ Physics: Proceedings of the Johns Hopkins
	Workshop on Current Pro\-blems in Particle Theory 12, Baltimore 1988''
	(G.\ Domokos and S.\ Kovesi-Domokos, Eds.)}}{World Scientific,
	Singapore, 1988.}

\end{thebibliography}
\end{document}